\documentclass[pdflatex,sn-mathphys-num, iicol]{sn-jnl}

\usepackage{graphicx}%
\usepackage{multirow}%
\usepackage{amsmath,amssymb,amsfonts}%
\usepackage{amsthm}%
\usepackage{mathrsfs}%
\usepackage[title]{appendix}%
\usepackage{xcolor}%
\usepackage{textcomp}%
\usepackage{manyfoot}%
\usepackage{booktabs}%
\usepackage{algorithm}%
\usepackage{algorithmicx}%
\usepackage{algpseudocode}%
\usepackage{listings}%
\usepackage{siunitx}
\usepackage{soul}
\usepackage{hyperref}

\theoremstyle{thmstyleone}%

\theoremstyle{thmstyletwo}%

\theoremstyle{thmstylethree}%

\begin{document}


\title[title]{Towards Neuromorphic Event-Based Sensing for High-Speed Multi-Spectral Classification and Tracking of Microparticles}

\author*[1,2]{\fnm{Joana} \sur{Teixeira}}\email{joana.m.teixeira@inesctec.pt}

\author[1,2]{\fnm{Tomás} \sur{Lopes}}

\author[1]{\fnm{Tiago} \sur{D. Ferreira}}
\author[1]{\fnm{Catarina} \sur{S.  Monteiro}}

\author[1,2]{\fnm{Pedro} \sur{A. S. Jorge}}
\author[1,2]{\fnm{Nuno} \sur{A. Silva}}

\affil[1]{\orgdiv{Centre for Applied Photonics}, \orgname{INESC TEC}, \orgaddress{\street{Rua do Campo Alegre 687}, \city{Porto}, \postcode{4169-007}, \state{Porto}, \country{Portugal}}}

\affil[2]{\orgdiv{Departmento de Física e Astronomia}, \orgname{Faculdade de Ciências, Universidade do Porto}, \orgaddress{\street{Rua do Campo Alegre s/n}, \city{Porto}, \postcode{4169-007}, \state{Porto}, \country{Portugal}}}

\abstract{Conventional image-based microfluidic systems face an inherent trade-off between throughput, imaging speed, and data bandwidth, limiting their ability to monitor high-velocity flows without significant motion blur or prohibitive data generation. Event-based sensing has emerged as a high-speed, low-power alternative, but has so far been largely restricted to tracking monodisperse, spherical particles. In this work, we introduce a microfluidic sensing platform that enables the simultaneous extraction of kinematic and spectral information from polydisperse microparticles using a neuromorphic imaging approach. By integrating a spatially multiplexed RGB filter mask with an asynchronous event-based sensor, spectral signature and motion are encoded directly at the sensing stage, eliminating the need for image reconstruction or learning-based inference. The system achieves sub-millisecond temporal resolution and maintains robust classification performance across a broad range of particle sizes and flow velocities, including under non-laminar conditions, reaching up to 82\% accuracy for classification of colored particles within the 0.08–0.18 mm range. The event-driven architecture reduces data bandwidth by $>$240× compared to conventional high-speed imaging, while sustaining an area throughput of 460 mm$^2$/s. By providing a computationally efficient and low-latency particle characterization, this framework paves the way for a scalable solution towards high-speed, label-free screening of heterogeneous analytes in clinical diagnostics and environmental monitoring.}


\maketitle

\section{Introduction}\label{sec1}
In the last decades, microfluidics have become an indispensable tool in a wide range of scientific and technological applications, ranging from drug discovery to single-cell analysis, to name a few examples \cite{Lombardo2021,Zhang2024}. A critical aspect of these microscale systems is the ability to accurately
track and characterize the behavior of particles or
cells that flow through microchannels. Currently, image-based microfluidic systems face inherent trade-offs between imaging speed, throughput, and data bandwidth. High-speed CMOS or CCD imaging is often required to capture fast dynamics, but may still lead to motion blur, reduced signal quality, and rapidly increasing data volumes that hinder real-time processing \cite{zhou2023computer}. As a result, increasing the throughput of the system, sometimes crucial for screening large populations, often requires sacrificing either imaging speed, sensitivity, or resolution. 

Several techniques have been proposed to overcome these limitations, in particular focusing on increasing the throughput of the systems and improving the resulting image quality. For instance, stroboscopic illumination has been successfully integrated into sheathless systems to provide blur-free analysis at throughputs exceeding 50,000 cells/s \cite{rane2017high}. By utilizing short, high-intensity light pulses, these systems reduce effective exposure times and increase the depth of field, though they require precise temporal synchronization and high-power light sources. Parallelization via multi-channel architectures or the integration of microlens arrays \cite{holzner2018optofluidic} has further pushed monitoring capacities, offering magnification and enumeration performance comparable to high-end microscopy across a wide field of view, at the cost of an increase in system complexity. More recently, the use of deep learning architectures has demonstrated the ability to enhance or reconstruct low-resolution images \cite{huang2022deep, siu2023optofluidic}, but introduces significant computational overhead and may compromise the fidelity of spectral and morphological information. Despite these advances, existing approaches remain fundamentally tied to frame-based imaging and its associated constraints.

These technical constraints have motivated the exploration of alternative imaging approaches, such as the use of event-based cameras \cite{howell2020high,willert2022event,zhang2022work}. In contrast to adaptations of frame-based architectures, event-based sensing (EVS), also known as neuromorphic vision sensors, may offer a fundamental shift in the imaging paradigm by discarding the concept of global frames altogether. By recording intensity changes asynchronously at the pixel level, EVS can be a powerful alternative for high-speed microfluidic analysis, benefiting from higher dynamic range, microsecond-scale temporal resolution, and promising lower memory and computing overhead \cite{lenero2018applications, cabriel2023event}.In fact, this technology has been demonstrated as a reliable approach for tracking and reconstruction of high-speed fluid flows with velocities up to \SI{1.7}{\meter\per\second} \cite{willert2022event,wang2020stereo, franceschelli2025assessment}. Additionally, the use of event-based cameras has been reported for capturing microparticles moving through microfluidic channels and classifying them according to their size, utilizing advanced neural network architectures \cite{tsilikas2024photonic, zhang2022work}. By processing the asynchronous event data directly, these deep learning approaches can extract morphological features and categorize the particle size with minimal latency. These results highlight the potential of event streams for high-dimensional characterization.

Despite these advances, the exploitation of spectral information from opaque analytes in event-based microfluidic architectures remains largely underexplored. In recent years, and in the context of microscopy, spectral and hyperspectral imaging have attracted increasing interest because wavelength-dependent reflectance and absorption can provide material- and chemistry-specific signatures that are not accessible from morphology alone \cite{zhang2023accumulation,banu2024hyperspectral, tran2024detection}. Event-based vision has also begun to be explored for spectral imaging recently, with active illumination schemes demonstrating that spectral information can be encoded into asynchronous event streams while reducing bandwidth requirements \cite{chen2026self}. However, to the best of our knowledge, the integration of event-based spectral sensing with continuously flowing samples in a microfluidic channel has not yet been demonstrated. Besides, previous event-based microfluidic studies have largely relied on monodisperse or well-defined particles, such as spherical beads with known geometries, or on learning-based classification pipelines. Such idealized conditions do not fully capture the morphological and optical complexity of real-world samples, where analytes often exhibit substantial heterogeneity in size, shape, and reflectance \cite{Mohammadimehr2024,Khan2026}.

In this work, we aim to move beyond these constraints by developing a neuromorphic microfluidic sensing platform capable of simultaneous tracking, sizing, and performing color classification of polydisperse fragments. By combining a spatially multiplexed RGB filter mask with an asynchronous event-based sensor, spectral identity and motion are encoded directly at the sensing stage, removing the need for training. We demonstrate real-time tracking, sizing, and color classification under non-laminar flow conditions, establishing a data-efficient and low-latency approach for high-throughput analysis of heterogeneous analytes, and paving the path for future multispectral analytical on-chip solutions.

\section{Methods}

This work combines a microfluidic flow platform, a spatially multiplexed optical filtering strategy, and an event-based processing pipeline to enable simultaneous spectral and kinematic characterization of microparticles. The following sections describe the experimental architecture of the system, including the microfluidic circuit, illumination geometry, and dual-camera imaging configuration, before detailing the calibration and signal-processing procedures used for color prediction, velocity estimation, and size extraction.

\subsection{System Architecture and Microfluidics} 

\begin{figure*}[h!]
    \centering
    \includegraphics[width=\linewidth]{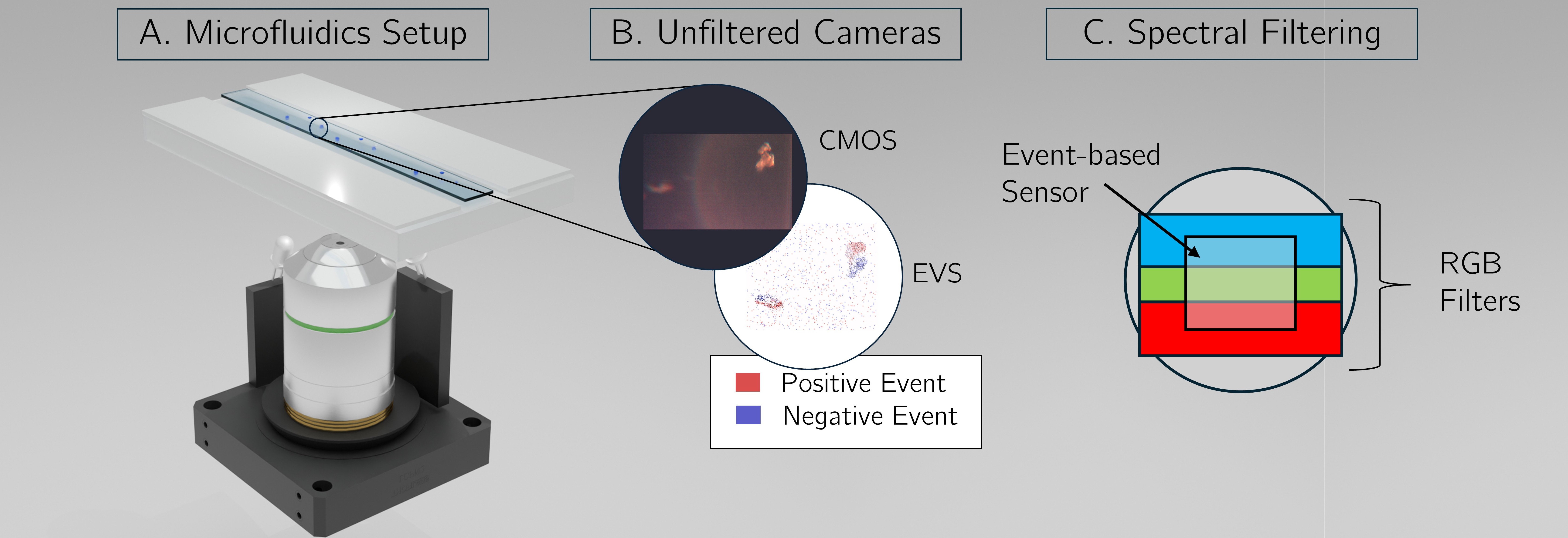}
    \caption{\textbf{Schematic representation of the experimental setup. (A)} The microparticle solution flows through the microfluidics channel and is illuminated with white LEDs. The reflected light is then collected by a 10x microscope objective. \textbf{(B)} The flow is imaged with both a conventional CMOS camera (top frame) and an event-based camera (bottom frame), with the later registering the movement of the particles through the activation of positive and negative events along its edges. \textbf{(C)} Different RGB filters are placed in front of non-overlapping regions of the event-based camera sensor, causing the passage of the particles to activate more pixels when their reflected spectrum matches the transmitted wavelengths of the filters.}
    \label{fig:setup}
\end{figure*}

The fluidic circuit is driven by a programmable syringe pump (NE-4002X Programmable 2-Channel, SyringeTWO), which controls the volumetric flow rate of the particle suspension. The suspension is delivered from a 3\,ml syringe through flexible tubing into a commercially available microfluidic channel (ibidi $\mu$-Slide, 0.2 Luer, glass bottom, channel dimensions: 50×5×0.2\,mm). The use of a glass-bottom slide ensures high optical transparency and compatibility with high-resolution microscopy. The channel is positioned in an inverted setup, with a 10x objective, and illumination in reflection is provided by a two-white-LED arrangement(see Figure \ref{fig:setup}A), enabling effective detection of opaque particles. By collecting backscattered light, the event-based sensor responds to the particles’ intrinsic color and surface reflectance, rather than solely to intensity occlusion. This configuration avoids the "shadow effect", where an event camera might only register the absence of light rather than the specific spectral characteristics of the particle. Additionally, the 10x objective was specifically selected to provide a large depth of focus, increasing the volume of the channel in which particles can be effectively detected and resolved. The light collected by the objective is directed through a beam splitter, which divides the optical signal between two distinct imaging paths.

To resolve individual particles with high precision, the light in each optical path is focused onto the respective sensors using achromatic lenses with a focal length of f=60\,mm. This configuration provides the necessary magnification to bridge the scale between the microfluidic channel and the sensor pixels. The system leverages two distinct sensing technologies to capture the dynamics of the flow: a frame-based camera (DCC1249C-HQ, Thorlabs), which provides a high-fidelity spatial reference for traditional intensity measurements, and a high-resolution event-based camera (IDS uEye XCP-E, 1280x720 px). By splitting the light path between these two sensors, the setup allows for a direct temporal and spatial correlation between traditional frame-based snapshots and the continuous, asynchronous data stream generated by the event-based architecture.

In the event-based imaging path, a multispectral filter assembly was positioned directly in front of the EVS, as shown in Figure \ref{fig:setup} C. This set is composed of three Lee photographic filters (in the colors Tokyo Blue, Primary Green, and Marius Red) specifically chosen for their minimal overlapping transmission across the visible spectrum (see Supplementary Material Fig. S5). To maintain a high signal-to-noise ratio and prevent "ghost events" triggered by ambient light, the region surrounding the filters is shielded with an opaque layer.

\subsection{Signal Processing and Calibration}
Event-based cameras operate on an asynchronous, per-pixel change-detection architecture. Unlike standard frame-based sensors that capture global snapshots, each pixel independently monitors incident radiance. By encoding intensity on a logarithmic scale, the sensor maintains an internal reference that is updated only when a predefined change threshold is exceeded. Each resulting event is defined by its spatial coordinates (x,y), a microsecond-resolution timestamp t, and a 1-bit polarity $p\in\left\{-1,+1 \right\}$, indicating whether the brightness increased or decreased. This data-driven nature ensures that regions of high dynamic activity generate dense data streams, while static scene components produce no output, drastically reducing redundancy. In this framework, an individual event $e_k=(x_k,y_k,t_k,p_k)$ is generated at pixel $\mathbf{x}_k=(x_k,y_k)$ at time $t_k$ whenever a change in the log-brightness L is detected since the previous event at that same location. This temporal change, $\Delta L$, is expressed as:
\begin{equation}
    \Delta L(\mathbf{x}_k, t_k) = L(\mathbf{x}_k, t_k) - L(\mathbf{x}_k, t_k - \Delta t_k)
\end{equation}
where $\Delta t_k$ represents the time interval elapsed since the last event recorded at that pixel. 

To enable processing with standard computer vision architectures, the asynchronous event stream was temporally discretized into frame-like representations via pixel-wise accumulation. In this work, discretization was performed according to specific acquisition frequencies, specifically 500 Hz, 1000 Hz, and 1500 Hz. To further reduce data dimensionality and enhance the signal-to-noise ratio, the events are spatially aggregated into 10×10 bins by summing the individual event contributions within each bin.

To identify the areas of the event-based images corresponding to each spectral channel, the microscope objective was illuminated with time-modulated LEDs emitting in the red, green, and blue bands. This meant that each LED highlighted the region corresponding to one of the filters, which we registered and saved as binary masks for subsequent analysis. 

On another note, the size of each image pixel was also calibrated against its corresponding sample size. For this, a square reticle was placed at the sample plane and imaged with both the CMOS camera and the event-based camera, the latter via modulated illumination. The pixel pitch was then estimated by fitting four straight lines to the sides of the imaged square. 

\subsubsection{Color Prediction}
\label{sec:color_prediction_process}
Starting with the binned frames generated as described in the previous section, the primary objective is to isolate signal from noise by leveraging the inherent spatial correlation of true event activity. In this implementation, only positive events were retained to minimize the computing time. However, the approach can be generalized to extract color predictions from both positive and negative events. To further suppress noise, a neighbor-count filtering strategy was applied to each binned frame. This leverages the fact that stochastic noise events typically occur in isolation, whereas true signal events exhibit local spatial clustering.

The process involves convolving the binary mask of active pixels with a 3×3 kernel to determine the local connectivity of each event. We utilize an 8-connectivity kernel:

K =$ \begin{bmatrix} 
1 & 1 & 1 \\ 
1 & 0 & 1 \\ 
1 & 1 & 1 
\end{bmatrix}$
\\

This kernel computes the sum of the active neighbors for each pixel, excluding the pixel itself. To ensure a high signal-to-noise ratio for subsequent spectral classification, we apply a thresholding operation where a pixel is retained only if its intensity exceeds a value of 1.0 and it possesses a minimum of n=1 neighbors. Any pixel failing to meet these criteria is considered an isolated noise event, and its value is set to zero. This neighborhood filtering significantly cleans the event trajectories associated with microparticles, ensuring that spectral predictions are based on dense clusters corresponding to physical particle transits, rather than background sensor fluctuations.

The spectral identity of a microparticle is determined by a maximum-likelihood decision based on the peak event activity observed within each filtered region. Rather than evaluating a single global frame or averaging activity over the entire transit duration, our algorithm performs an independent temporal search for each spectral band. For every frame $k$ in the discretized sequence, an instantaneous score $s_{c,k}$ is calculated for each color $c \in \{R, G, B\}$ by summing the filtered event intensities within the respective binned masks ($M_c$):

\begin{equation}
s_{c,k} = \sum_{(x,y) \in M_c} F_{k, \text{filtered}}(x,y)
\end{equation}

To account for the fact that a particle may reach peak sensor response at different times for each spectral channel, owing to its trajectory across the physical sensor, the system identifies the peak score for each color independently across the entire frame sequence:
\begin{equation}
S_{c, \text{peak}} = \max_{k} (s_{c,k})
\end{equation}

The final classification is then determined by comparing these three independent maxima. The predicted color corresponds to the spectral region that yielded the highest integrated event peak during the transit:
\begin{equation}
\text{Prediction} = \operatorname{argmax}_{c \in \{R, G, B\}} (S_{c, \text{peak}})
\end{equation}

This "Best-of-Frames" strategy per color ensures that the classification is based on the highest quality signal for each band, regardless of whether the particle was perfectly centered in all regions at the same moment. Furthermore, retaining the scores for the non-dominant regions ($S_{\text{second}}$) enables a secondary likelihood analysis. This is particularly critical for characterizing particles with complex or overlapping spectral profiles, such as those exhibiting dual-peak signatures. 

\subsubsection{Velocity Estimation}
\label{sec:velocity_method}
Following the spatial denoising and binning of the event stream, the kinematic profile of each microparticle is extracted through a multi-step tracking algorithm. This process involves the determination of the particle’s center of mass in each discretized frame and the subsequent linking of these coordinates into a continuous temporal trajectory.

For each filtered frame $F_{k,filtered}$, the spatial centroid $(x_c,y_c)$ is calculated to represent the particle's instantaneous position. To ensure robustness against noise, a minimum total energy threshold is applied. If the sum of binned event intensities fails to meet this threshold, the frame is discarded from the tracking analysis. The centroid is calculated as:
\begin{equation}
x_c = \frac{\sum (x \cdot w(x,y))}{\sum w(x,y)}, \quad y_c = \frac{\sum (y \cdot w(x,y))}{\sum w(x,y)}
\end{equation}
where $w(x,y)$ represents the event intensity at coordinates $(x,y)$.

Particle trajectories were reconstructed using a proximity-based linking algorithm. A "Track" object was initialized for each new detection, maintaining a history of spatial coordinates and timestamps. For each subsequent frame, detections were associated with existing tracks by identifying the nearest candidate within a maximum spatial jump threshold. To account for transient signal loss, which is often caused by particles passing through low-contrast regions or moving out of focus, the tracker allows for a predefined number of "missed" frames before a track is officially closed. Note that for a certain frame sequence, several tracks can be identified due to noisy regions that have bypassed the noise-filtering pipeline. Under these circumstances, the track that spans the highest number of frames is considered, as the noise tracks typically persisted for only 1-2 frames.

Once a trajectory is established, the instantaneous velocity components $(v_x,v_y)$ are derived from the displacement between consecutive centroids. Given the calibrated pixel scale $s(mm/px)$ and the frame integration time $\Delta t$(s), the velocity is calculated as:
\begin{equation}
v_x = \frac{\Delta x \cdot s \cdot b}{\Delta t}, \quad v_y = \frac{\Delta y \cdot s \cdot b}{\Delta t}
\end{equation}
where b is the binning factor. The overall mean speed $v_{mean}$ for a particle transit is then determined by averaging the magnitude of the velocity vector $v=v_x^2+v_y^2$ across the entire track history. This method allows for kinematic profiling even for irregular particles, providing a significant advantage over frame-based systems that may suffer from motion-blur-induced centroid shifts.

\subsubsection{Size Estimation}
\label{sec:size_method}
While the kinematic tracking relies exclusively on positive polarity events, the size estimation incorporates the absolute intensity of both positive and negative polarities, thereby capturing the full physical boundary of the microparticle. To facilitate robust shape estimation, the event stream is processed through a sequence of morphological steps. 

Following spatial binning, the event-based data, which can be sparse or "disconnected" depending on the particle's local contrast, was subjected to a binary closing operation, using a disk of radius 2 (scikit-image). This fills internal gaps within the particle's event cluster without significantly altering its external dimensions. Subsequently, spatial consistency was enforced by applying an 8 connectivity criterion and retaining only connected components exceeding a total area of 20 pixels, thereby reducing isolated noise artifacts.

To account for the irregular and size variability of the particles, their morphology is modeled as an equivalent ellipse by calculating the eigenvectors of a weighted spatial covariance matrix. To prioritize the dense core of the particle over potentially noisy peripheral events, we employ a dual-weighting scheme where each pixel $i$ in the mask is assigned a weight $w_i$ combining its local event intensity $w_{int}$ with a Gaussian center-weighting term $w_{center}$:
\begin{equation}
w_i = w_{\text{int}} \cdot \exp\left(-\frac{|\mathbf{P}_i - \mathbf{P}_{\text{COM}}|^2}{2\sigma_c^2}\right)
\end{equation}
where $\mathbf{P}_i$ is the pixel coordinate, $\mathbf{P}_{\text{COM}}$ is the weighted center of mass, and $\sigma_c$=10 pixels. The weighted covariance matrix $\mathbf{C}$ is then defined as:
\begin{equation}
\mathbf{C} = \frac{\sum w_i (\mathbf{P}_i - \bar{\mathbf{P}})(\mathbf{P}_i - \bar{\mathbf{P}})^T}{\sum w_i}
\end{equation}
The eigenvalues $\lambda_1, \lambda_2$ of $\mathbf{C}$ provide the variance along the principal axes. The semi-major (a) and semi-minor (b) axes are derived using a scaling factor of 2.0 (representing a $2\sigma$ boundary):
\begin{equation}
a = 2\sqrt{\lambda_1}, \quad b = 2\sqrt{\lambda_2}
\end{equation}

By multiplying the binned results by the calibration factor, we obtain a morphological profile that accounts for the irregular nature of the samples while remaining consistent with conventional CMOS "ground truth" measurements.

\subsection{Sample description}
To validate the robustness of our neuromorphic spectral-kinematic platform, we aimed to move beyond the idealized constraints of monodisperse, spherical microparticles. In real-world applications, ranging from clinical diagnostics to environmental monitoring, analytes are rarely uniform \cite{Khan2026}. Therefore, we developed a sample preparation method designed to simulate the morphological and spectral diversity inherent in complex, non-idealized samples. In specific, polychromatic microparticles were created via mechanical abrasion of colored thermoplastic polymers using an electric rotary abrasive tool. This process yielded a population of particles characterized by extreme heterogeneity in both morphology and size distribution, with irregular geometries and sizes ranging from a few to hundreds of micrometers. The effective size range utilized in this study was naturally constrained by the microfluidic system, with the channel interfaces acting as a functional filter, eliminating the largest fragments that could otherwise cause blockages. Conversely, some of the smallest fractions might have been excluded due to buoyancy effects and limited transport within the flow.

Before injection, the particles were suspended in deionized water and subjected to 10 minutes of ultrasonication to promote deagglomeration. To mitigate the risk of plastic particles adhering to the hydrophobic surfaces of the fluidic delivery system, the syringes were first preconditioned with a dilute surfactant solution. Following preconditioning, the syringes were flushed and refilled with the prepared microparticle suspension.

By utilizing these heterogeneous particles, this study aims to serve as a proof-of-concept for more demanding, real-world cases. Whether monitoring microplastic pollution in environmental samples or identifying biological cells in clinical flows, the ability to accurately classify and size non-spherical, heterogeneous analytes is a critical requirement that conventional frame-based systems struggle to meet due to motion blur and fixed exposure constraints. It should be noted that, as a consequence of the procedure used to generate the microparticles, a small degree of cross-contamination between color populations cannot be completely excluded.

\section{Results and Discussion}
The performance of the neuromorphic spectral-kinematic pipeline was evaluated using a heterogeneous population of red, green, and blue microplastic fragments. Experiments were conducted under two flow conditions, with the syringe pump set to nominal flow rates of 1.334 ml/min and 3.000 ml/min. These rates were selected to test the sensor’s ability to resolve both moderate and high-velocity transits. However, the complexity of the experimental system, which included 3D-printed couplers between the syringe, tubing, and the $\mu$-slide, introduced significant deviations from idealized flow. The presence of residual air and water bubbles within the system, combined with the irregular geometries of the particles, resulted in non-laminar flow conditions. These factors, alongside the stochastic rotation and varied buoyancy of the particles, created a dynamically complex environment providing rigorous test conditions for real-time tracking and spectral classification.

\subsection{Kinematic Profiling: Size and Velocity}

\begin{figure*}[h!]
    \centering
    \includegraphics[width=\linewidth]{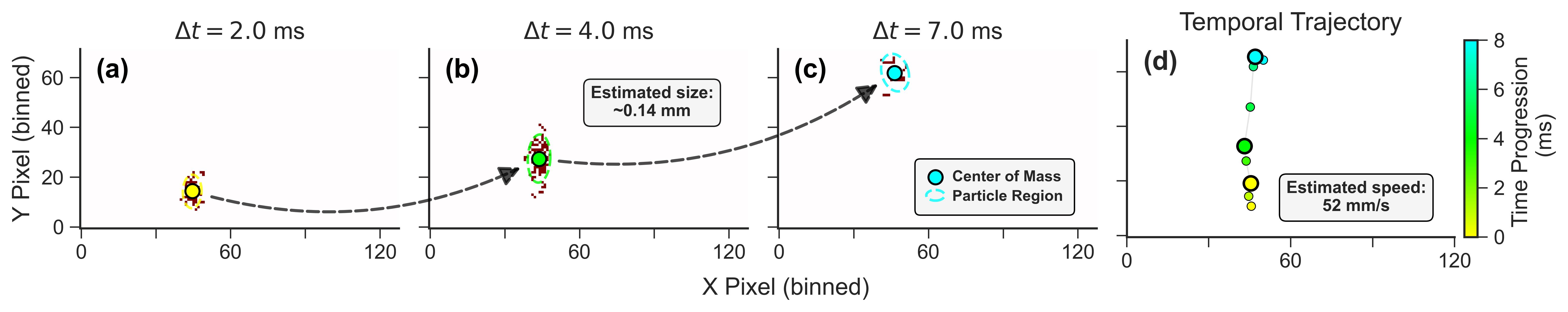}
    \includegraphics[width=\linewidth]{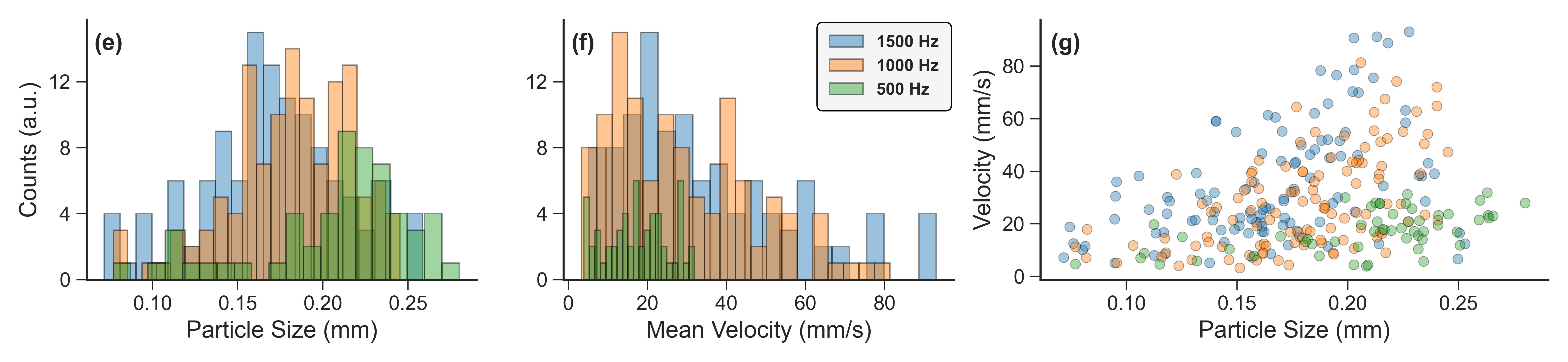}
    \caption{\textbf{Representation of the particle identification, tracking and sizing estimation. (a)–(c)} Event-based frames acquired at 1\,kHz after binning and processing, showing the particle at different positions along the channel. Pixels shown in red mark the events detected for each frame. The center of mass indicates the estimated particle position used for motion estimation. The dashed ellipses represent the detected particle region and are used to estimate the particle size from the ellipse's minor axis. \textbf{(d)} Reconstructed particle trajectory across the captured frames, with color indicating temporal progression. \textbf{(e)} Distribution of the estimated particle sizes obtained from the ellipse fits, for different frame construction frequencies. \textbf{(f)} Distribution of the estimated particle velocities across tracked segments. \textbf{(g)} Relationship between particle size and velocity for the detected particles.}
    \label{fig:tracking}
\end{figure*}

The primary advantage of the event-based approach in microfluidics is the decoupling of temporal resolution from data bandwidth, allowing for microsecond-precision tracking without the prohibitive storage requirements associated with high-speed frame-based cameras. Using the centroid tracking and "best run" selection logic described in Sections \ref{sec:velocity_method} and \ref{sec:size_method}, we successfully resolved the trajectories of particles across a wide size and velocity range. For these tests, events were integrated into temporal windows corresponding to acquisition rates of  500, 1000 and 1500\,Hz. Figure \ref{fig:tracking} presents an example of processed frames for a detected particle, where the estimated center of mass marks its position along the trajectory and a fitted ellipse outlines the particle boundaries.

The estimated particle size distribution (Fig. \ref{fig:tracking}e) shows that most particles fall within the range of approximately 0.05–0.30 mm. Particle size is estimated from the minor axis of the fitted ellipse, which provides a more stable approximation of the particle diameter. The particles observed in the experiments are irregular and can rotate due to flow turbulence, leading to frame-to-frame variations in the apparent contour of the particle. In addition, the major axis of the ellipse can be artificially enlarged by the spatial spreading of detected events along the direction of motion, particularly along the dominant flow axis. This effect results from the finite temporal integration window and the motion of the particle across the sensor during the acquisition window. As a consequence, the major axis may overestimate the true particle dimension, whereas the minor axis remains less affected by this motion-induced distortion. For this reason, the minor axis was adopted as a more robust estimator of particle size. In the supplementary material, Section S1, a comparison between the size estimation obtained from the RGB camera and the event-based camera is presented, demonstrating close agreement of the two cases. Additionally, it is shown that detecting the particle in or out-of-focus produces similar estimations of the particle size in the case of event-based measurements. 

The velocity distribution shown in Fig. \ref{fig:tracking}(f) indicates that most detected particles travel at speeds between approximately 3 and 80\,mm/s. Although the syringe pump was operated at nominal flow rates of 1,334 mL/min and 3,000\,mL/min, the presence of air bubbles and transient disturbances introduced deviations from ideal laminar flow conditions. For this reason, these flux values provide only an estimation of the order of magnitude of the velocities and should not be regarded as ground truth values. In fact, for the estimated particle sizes, the measured velocities are coherent with the range expected for the two flux conditions. A comparison between the measured particle velocities and the theoretical values expected from the nominal flow rates is presented in Fig. S3 in the Supplementary Material. 

While the estimations of both velocity and size show similar profiles for the frames registered at 1000 and 1500\,Hz, the measurements obtained at 500\,Hz exhibit noticeable deviations. At the lower acquisition frequency, the system is unable to adequately capture the highest particle velocities present in the flow, leading to an under-representation of fast-moving particles in the measured velocity distribution. In addition, the estimated particle sizes tend to appear larger when using the 500\,Hz acquisition rate. This behavior arises from the reduced temporal sampling, as fewer frames are available to capture the particle trajectory, leading to larger accumulated event regions in each frame. These enlarged regions may span multiple color filter areas and merge information from different spatial locations of the particle along its motion. Consequently, the segmentation step becomes less effective at separating the instantaneous particle positions, which can artificially increase the apparent particle size and reduce the accuracy of the velocity estimation.

These limitations at lower frame rates highlight the significant advantages of the event-based sensing approach. The high temporal resolution enables fast particle transits, while the flexibility of the event representation allows the processing parameters to be adapted during post-processing depending on the acquisition conditions. This adaptability makes it possible to optimize the tracking and sizing procedures according to the flow regime and particle dynamics observed in each experiment.

The relationship between particle size and velocity is illustrated in Fig. \ref{fig:tracking}(g), where a weak positive correlation is observed. This trend is consistent with the parabolic velocity profile expected in pressure-driven microchannel flow, in which the fluid velocity is larger at the center of the channel and decreases toward the channel walls. The observed correlation may therefore reflect a bias in the spatial distribution of particles across the channel, with larger particles more likely to be detected in regions of higher flow velocity.

It should also be noted that the detected particles exhibit irregular shapes and rotational motion as they travel through the channel. This behavior can introduce variations in the apparent particle contour across frames, thereby influencing the ellipse-fitting procedure used for size estimation. Despite these effects, the tracking method provides consistent estimates of both particle size and velocity, enabling a meaningful statistical characterization of the particle population within the observed flow.

\subsection{Color Prediction}
\begin{figure*}[h!]
    \centering
    \includegraphics[width=\linewidth]{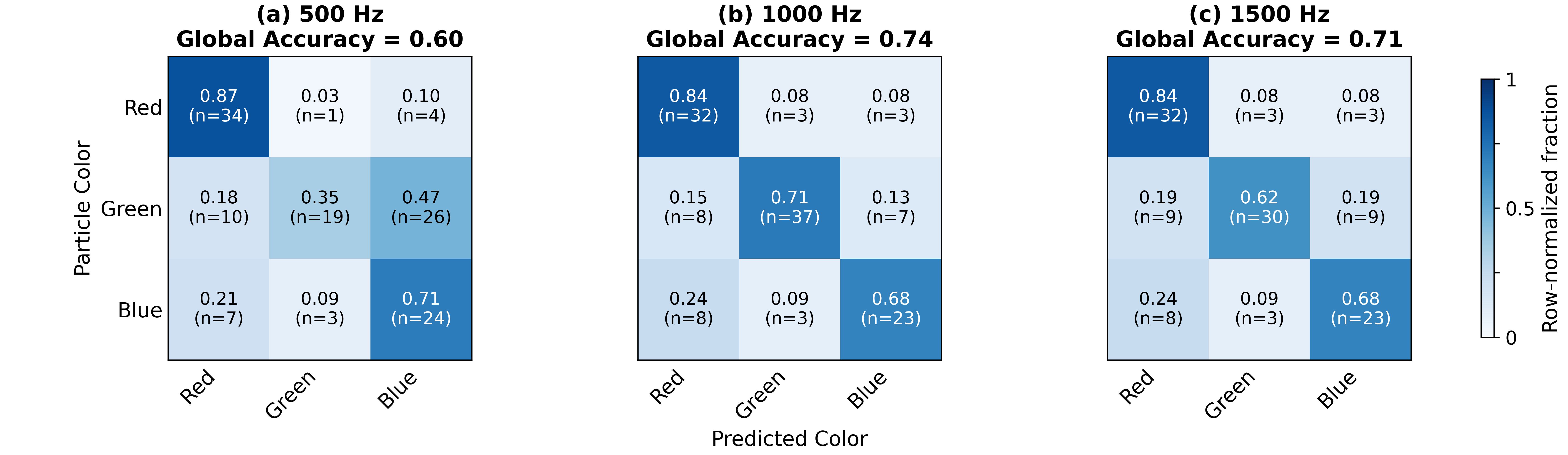}
    \caption{\textbf{Color classification performance at different frame construction frequencies. (a)–(c)} Row-normalized confusion matrices for the classification of particle color using the event-based sensing pipeline at frame construction frequencies of 500\,Hz, 1000\,Hz, and 1500\,Hz, respectively. Each matrix compares the predicted particle color against the ground truth label, with the color scale indicating the normalized fraction of predictions for each true class. The values inside each cell correspond to the normalized fraction of classifications, with the number of samples used for each entry indicated in parentheses. The overall classification accuracy for each frequency is reported in the title of each panel.}
    \label{fig:CM_all_particles}
\end{figure*}

The spectral identification capability of the proposed platform was evaluated by applying the color prediction algorithm described in Section \ref{sec:color_prediction_process} to the tracked particle events. In this framework, the spectral signature of each particle is inferred from the relative event activity generated within the spatial regions corresponding to the red, green, and blue filters. By extracting the peak event activity independently for each spectral band during the particle transit, the algorithm determines the most likely color class through a maximum-likelihood comparison of the resulting scores.

Figure \ref{fig:CM_all_particles} summarizes the color classification performance of the proposed event-based spectral pipeline for frame construction frequencies of 500, 1000, and 1500\,Hz. The predictions were obtained using the procedure described in Section \ref{sec:color_prediction_process}, in which the filtered event activity is evaluated independently in the spatial regions associated with the red, green, and blue filters, and the final class is assigned according to the maximum peak response during the particle transit. Overall, the results indicate that the intermediate and higher frame construction frequencies provide the most reliable spectral discrimination, with global accuracies of 74\% and 71\% at 1000 and 1500\,Hz, respectively, compared with 60\% at 500\,Hz.

Color classification performance at low acquisition frequency is limited by temporal undersampling. At 500\,Hz, the extended integration window allows particles to undergo significant displacement within a single frame, causing events generated at different positions along the trajectory to be accumulated together. Consequently, the measured signal reflects a spatially averaged response that partially mixes contributions from multiple spectral regions, thereby degrading class separability. This effect is most pronounced for intermediate spectral signatures, such as green, which are inherently more sensitive to overlap with neighboring channels.

The comparison between 1000 and 1500\,Hz suggests that simply increasing the frame construction frequency does not necessarily yield a proportional gain in classification performance. Although shorter integration times reduce color mixing, they also decrease the total number of events accumulated per frame, which weakens the particle signal. This trade-off explains why 1000\,Hz provides the best overall performance in the present experiments, while 1500\,Hz yields a similar but slightly lower accuracy. These results highlight the existence of an optimal operating regime that arises from a balance between temporal resolution and event density. In principle, the performance at higher frame construction frequencies could be further improved by increasing the illumination intensity, thereby enhancing the generated event signal without sacrificing the temporal discrimination needed to prevent cross-filter contamination.

Considering that 1000\,Hz seems to provide the best classification performance for the present datasets, frames constructed under these conditions were used for the subsequent analysis of the influence of particle size and velocity on spectral prediction. This analysis assesses whether the classification performance remains stable across varying particle morphologies and transit dynamics, or whether specific ranges of size and speed introduce additional ambiguity in the spectral response.

\begin{figure*}[h!]
    \centering
    \includegraphics[width=\linewidth]{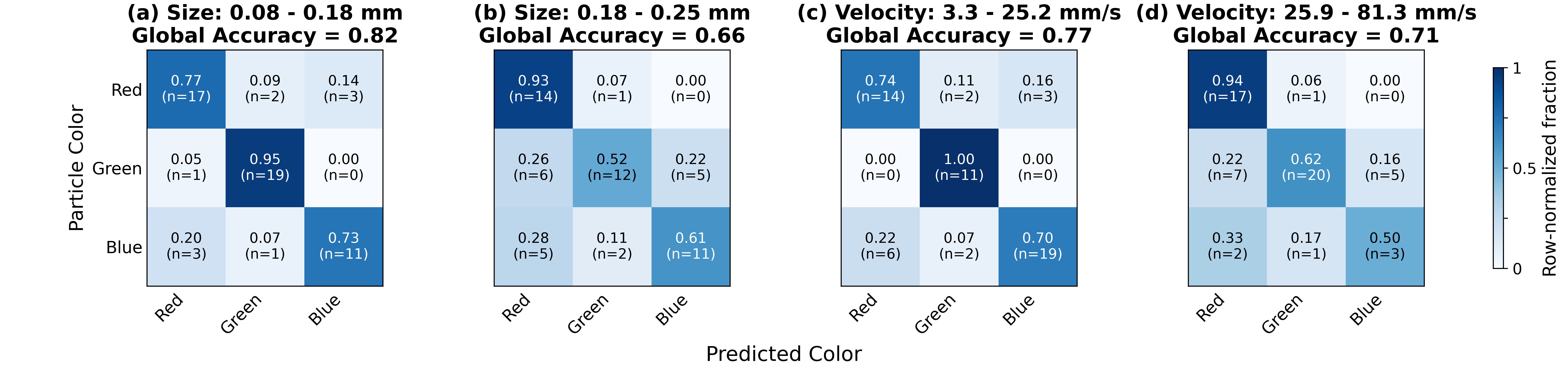}
    \caption{\textbf{Color classification performance for frames constructed at 1000\,Hz, according to different particle sizes and velocities. (a),(b)} Row-normalized confusion matrices for the classification of particle color using the event-based sensing pipeline at frame construction frequency 1000\,Hz according to the estimated particle size.  \textbf{(c),(d)} Row-normalized confusion matrices for the classification of particle color using the event-based sensing pipeline at frame construction frequency 1000\,Hz according to the particle velocity. Each matrix compares the predicted particle color against the ground truth label, with the color scale indicating the normalized fraction of predictions for each true class. The values inside each cell correspond to the normalized fraction of classifications, with the number of samples used for each entry indicated in parentheses. The overall classification accuracy for each group is reported in the title of each panel.}
    \label{fig:CM_persize_pervelocity}
\end{figure*}

These results are summarized in Fig. \ref{fig:CM_persize_pervelocity}. When considering the particle size intervals, a noticeable decrease in classification performance is observed for larger particles. While particles in the smaller size range (0.08–0.18\,mm) achieve a global accuracy of approximately 82\%, the accuracy decreases to 66\% for particles in the larger interval (0.18–0.25\,mm). This behavior can be attributed, in part, to the spatial extent of the event clusters generated by larger particles. Because the color filters are arranged as adjacent regions on the sensor, a particle with a larger apparent footprint is more likely to simultaneously activate pixels belonging to multiple filter areas, particularly when it passes close to the boundary between two spectral regions. Under these circumstances, the resulting event activity may produce comparable scores in more than one color mask, which increases the probability of misclassification. This can explain some of the confusion between Red and Green predictions as well as Green and Blue predictions, which doesn't happen for the case of the smaller particles. To address this limitation, Section S3 of the Supplementary Material introduces a more complex classification approach, which leverages a differential salience metric to decouple spectral identity from particle size. While the 'best-of-frames' logic used in this study is optimized for the high-throughput requirements of real-time flow, the salience-based method provides a high-precision alternative for characterizing larger, irregularly shaped analytes. By normalizing local event density relative to the specific background of each spectral region, this approach demonstrates how the platform can be generalized for more demanding morphological scenarios, though its reliance on global frame normalization makes it better suited for controlled calibration or low-density flows where signal masking between neighboring particles is minimized.

\begin{figure*}[h!]
    \centering
    \includegraphics[width=\linewidth]{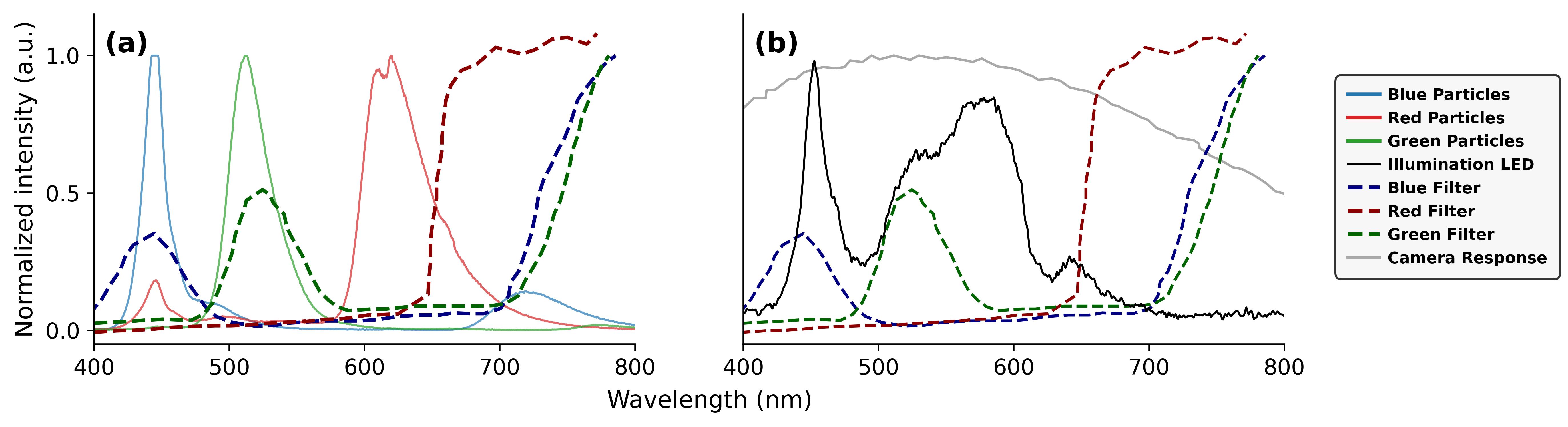}
    \caption{\textbf{Comparison between particle reflection spectra and transmission profiles of the RGB filters used in the event-based sensor. (a)} Normalized reflection spectra of the blue, red, and green microparticles (solid lines) overlaid with the transmission profiles of the corresponding blue, red, and green optical filters (dashed lines). This panel demonstrates the spectral alignment between each particle type and its respective filter. \textbf{(b)} Comparative transmission profiles of the three photography filters (dashed lines) alongside the emission spectrum of the illumination LED (solid black line) and the camera's quantum efficiency (solid grey line). This panel highlights the distinct spectral windows of the filters and their relationship to the light source, which enables the event-based system to infer particle color by identifying the filter region producing the strongest response during transit. The particle spectra were normalized to each maximum and obtained from reference measurements, while the filter transmission profiles correspond to standard specifications from the filter catalog.}
    \label{fig:filter_vs_particle}
\end{figure*}

To better understand the remaining classification errors, Figure \ref{fig:filter_vs_particle} compares the reflection spectra of the particles with the transmission profiles of the RGB filters. While the filters were selected to minimize spectral overlap, some cross-sensitivity remains due to the broad spectral response of the particles. In particular, the red and blue particle spectra exhibit partial overlap in the long-wavelength region of the blue response and the short-wavelength tail of the red response. Because the color prediction relies on identifying the filter region that produces the strongest event activity, particles with overlapping spectral components may generate detectable responses in more than one filter region. This effect likely contributes to the residual confusion between Red and Blue predictions observed in the confusion matrices. It should also be noted that the effective response measured by the event-based sensor is influenced by the spectral sensitivity of the camera and by the emission spectrum of the illumination source, which jointly determine the effective spectral response of the system and thus the relative signal amplitudes across spectral bands. For particles with spectral peaks of similar intensities across multiple bands, this presents a significant challenge for a simple maximum-likelihood classification. This is particularly evident in samples that exhibit multi-modal spectral profiles, where a single particle may trigger substantial event activity in two adjacent filter regions. While customized filters and illumination would surely enable improved performance, here, we focus on a proof-of-concept with off-the-shelf components. Still, we propose two distinct strategies for enhancing classification accuracy in complex scenarios. First, at the hardware level, the illumination source can be optimized by utilizing multi-wavelength LEDs tailored to the specific application, in order to better match the filter responses and reduce spectral cross-talk. Alternatively, at the algorithmic level, for applications requiring a more versatile, source-independent approach, we can implement a multi-rank prediction logic. Instead of relying on a single winning class, the event stream is used to generate the top two color predictions, which are then scored against one another to identify dual-peak signatures. To further explore the limits of the system's spectral sensitivity, a separate experiment using a specialized population of particles with both significant green and blue peaks was performed. This discussion, using the top-two color predictions, is presented in section S2 of the supplementary material.

A similar, though less notorious, trend is observed when analyzing the results as a function of particle velocity. The lower velocity interval (3.3–25.2\,mm/s) yields a higher global accuracy of approximately 77\%, whereas the accuracy decreases slightly to 71\% for faster particles (25.9–81.3\,mm/s). This behavior can be attributed to the increased displacement of fast-moving particles within the same temporal integration window. As particle velocity increases, the spatial extent of the accumulated event signal grows, raising the likelihood that events originating from different spectral regions are combined within the same frame. This effect is analogous to the behavior previously discussed for lower acquisition frequencies, where longer integration windows increase the probability of spatial mixing between filter regions. Consequently, although the 1000\,Hz frame construction frequency provides a good balance between event density and temporal resolution, very high particle velocities can still introduce some degree of spectral ambiguity due to the rapid displacement of the particle across the filtered regions of the sensor.

\subsection{Comparative Analysis: EVS vs. Conventional Imaging}

\begin{figure*}[h!]
    \centering
    \includegraphics[width=\linewidth]{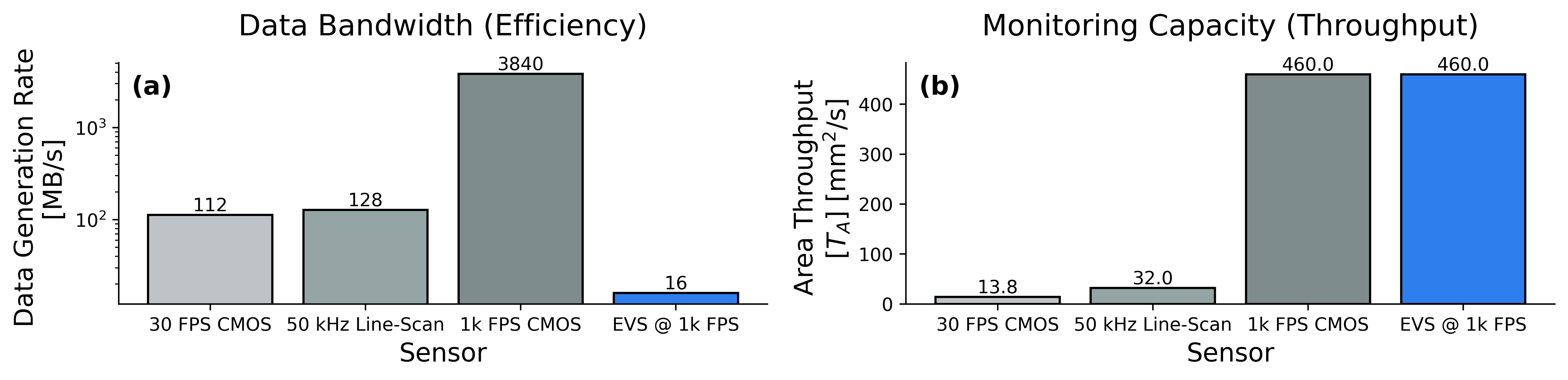}
    \caption{\textbf{Performance comparison of the proposed event-based spectral imaging system against conventional image-based architectures.} \textbf{(A) Data Bandwidth (Efficiency):} Comparison of data generation rates (MB/s) on a logarithmic scale. The EVS @1k FPS configuration achieves a $>$240× reduction in data volume compared to a 1k FPS CMOS baseline. Even when compared to a standard 30 FPS CMOS camera, the asynchronous nature of the EVS results in a significantly lower data footprint by eliminating background redundancy. \textbf{(B) Monitoring Capacity:} Comparison of the Area Throughput, representing the total fluidic area scanned per second. While the 30 FPS CMOS and 50 kHz line-scan cameras are limited in their capacity to monitor high-velocity flows, the EVS maintains a throughput of 460 mm$^2$/s while utilizing the sensor's full resolution.}
    \label{fig:comparison_stats}
\end{figure*}

To further illustrate the limitations of conventional frame-based imaging under our experimental conditions, supplementary videos acquired with a conventional frame-based camera (DCC1240, Thorlabs) operating at its maximum frame rate (17.4 fps) are provided \href{https://drive.google.com/drive/folders/181SGZwdnkK0GKvcQy3aMci5qp4c3WEis?usp=sharing}{here}. Note that these versions of the videos have been slowed to improve visualization. The recordings clearly exhibit significant motion blur, particularly for faster-moving particles, which often appear only as faint streaks. While slower particles can remain visible in multiple frames, faster particles cannot be reliably characterized in terms of size or speed, highlighting the fundamental limitations of low-frame-rate acquisition in high-throughput microfluidic flows.

To quantify the operational efficiency of the neuromorphic approach, we conducted a comparative data footprint analysis between our event-based sensor and a conventional frame-based color camera. The CMOS camera, with a 1280$\times$1024 resolution, captured approximately 3,841 KB of raw data per frame. To achieve the temporal resolution necessary for tracking high-velocity particles without significant motion blur, equivalent to an acquisition rate of 1,000 frames per second, such a system would generate a data stream of approximately 3.84\,GB/s. This staggering volume of data consists largely of redundant background information, as the sensor must sample every pixel regardless of whether a particle is present in the field of view.

In contrast, our 1280$\times$720 EVS system recorded a continuous 100-second microfluidic stream with a final compressed file size of 1,601,701\,KB ($\approx$ 1.6\,GB). This results in an average data rate of approximately 16\,MB/s, representing a $>$240× reduction in storage requirements compared to the 1,000 FPS CMOS baseline. It is noteworthy that these values were obtained under high-sensitivity bias configurations. We intentionally tuned the neuromorphic sensor to maximize event generation, ensuring that faint or out-of-focus particles, which might otherwise fall below the quantization threshold of the camera, triggered sufficient activity for detection and tracking. Therefore, for applications with a restricted depth of field, it is possible to decrease the data rate even further.

Beyond storage savings, the neuromorphic approach offers superior flexibility in post-processing. Unlike conventional frame-based cameras, which are fundamentally locked into a fixed frame rate at the time of acquisition, the asynchronous nature of EVS data allows the user to "slice" the event stream into arbitrary temporal bins during post-processing. This enables microsecond-precision tracking for high-speed kinematics while maintaining a total storage footprint lower than that of a standard 30 FPS camera. This paradigm shift from frame-based to event-based acquisition effectively decouples temporal resolution from data volume, allowing for high-throughput analysis without the prohibitive computational and storage overhead typically associated with high-speed microfluidic imaging.

A different alternative relies on the use of snapshot multispectral cameras, which utilize a mosaic of Fabry-Pérot filters or complex light-splitting optics to capture spectral data across a 2D plane in a single exposure \cite{gao2015optical}. While this provides high spectral resolution, it is fundamentally limited by the global shutter's frame rate, typically capped at 100 Hz for high-resolution sensors \cite{hagen2013review}. In high-speed microfluidics, this results in significant motion blur and a "blind time" between frames, during which fast-moving particles can bypass the field of view entirely. Furthermore, the data overhead is extreme: every pixel in the 3D hypercube is recorded at every time step, regardless of whether a particle occupies that voxel.

Moreover, line-scan cameras are often associated with high temporal resolution, as they only read out a single row of pixels ($1×N_x$) at each acquisition step. However, this is a misleading metric for particle tracking. To reconstruct a coherent morphological image and follow a particle’s trajectory across a flow cell, a system must capture a vertical sequence of lines equivalent to the sensor height ($N_y$). Therefore, the effective frame rate ($f_{eff}$) is not the line rate ($L$), but rather $f_{eff}=L/N_y$. For a standard line sensor operating at a 50\,kHz line rate \cite{liu2018novel}, to achieve y-axis resolution comparable to that used in our EVS, the effective frame rate would drop to approximately 70\,Hz, or 700\,Hz if we consider the 10x10 spatial binning step applied in the processing steps.

\subsubsection{Throughput Analysis: Area Scanning Rate (cm$^2$/s)}

To standardize the comparison across different imaging architectures, we defined the area throughput ($T_A$) as the total fluidic area monitored per second without the onset of significant motion blur. Utilizing our calibration scale of 0.71 $\mu$m/px, the 1280×720 sensing area corresponds to a physical field of view of approximately 0.91\,mm×0.51\,mm, or a total area of 0.46\,mm$^2$. If we consider a post-processing frame rate of 1000 fps, then this results in a $T_A$ of 460mm$^2$mm/s.

For a standard CMOS camera operating at a typical 30 FPS, and assuming the same field of view, the resulting area throughput is limited to 13.8\,mm$^2$/s. Although increasing the frame rate could theoretically improve this, the proportional increase in data bandwidth and decrease in exposure time often render size estimation unreliable for particles moving at high velocities.  Similarly, a high-end line-scan camera operating at a line rate of 50 kHz suffers the 2D reconstruction penalty. As the line rate must be divided by the 720 vertical pixels, it results in a corresponding $T_A$ of 32\,mm$^2$/s. These results have been summarized in Figure \ref{fig:comparison_stats}.

\section{Conclusion}
We have demonstrated a robust neuromorphic spectral-kinematic platform capable of simultaneously sizing, tracking, and color-classifying highly heterogeneous, non-idealized microparticles. By moving beyond the constraints of traditional frame-based imaging, our system successfully resolves the dynamics of irregularly shaped particles under non-laminar flow regimes, where motion blur and fixed exposure times typically constrain performance.

The integration of a spatially-multiplexed RGB filter mask with an asynchronous event-based architecture enables microsecond-precision kinematic profiling alongside high-fidelity spectral identification. For the flow conditions studied in this work, a frame-construction frequency of 1000\,Hz provides an optimal balance between temporal resolutions and signal density, achieving a global classification accuracy of 74\%. Furthermore, we have shown that while morphological irregularity and high velocities can introduce spectral ambiguity, the implementation of a "Best-of-Frames" strategy and multi-rank likelihood analysis allows the system to resolve complex, dual-peak spectral signatures that would be lost in standard single-threshold classifiers.

Moreover, by eliminating background redundancy, the event-based pipeline achieves a $>$240$\times$ reduction in data bandwidth relative to a 1000 FPS CMOS baseline, while maintaining an area throughput of 460 mm$^2$/s. This decoupling of temporal resolution from data volume enables the monitoring of high-velocity transits without the prohibitive storage and computational overhead associated with high-speed imaging or hypercube-based multispectral imaging systems.

Ultimately, this work serves as a proof-of-concept for the next generation of flow cytometry and environmental monitoring tools. Whether identifying biological cells in clinical diagnostics or monitoring microplastic pollution in environmental samples, the ability of this neuromorphic platform to handle polydisperse, irregularly shaped analytes makes it a versatile solution for real-world sensing applications. Importantly, the results presented here were achieved under deliberately non-ideal, application-relevant conditions, using heterogeneous particle populations, non-laminar flow dynamics, broad white illumination, and commercially available off-the-shelf RGB photographic filters rather than an optimized spectral architecture. The fact that both spectral classification and kinematic profiling were demonstrated under these constraints highlights the robustness of the event-based approach and suggests that the reported performance should be regarded as a conservative baseline. For applications requiring higher classification accuracy, reduced spectral cross-talk, or operation in critical sensing scenarios, the same architecture could be substantially improved through dedicated optical design, including application-specific filter sets with sharper and better-separated transmission bands, tailored multi-wavelength illumination sources matched to the target analytes, and optimized filter geometries on the sensor. Future work will therefore focus on moving from this low-complexity demonstrator toward purpose-built multispectral event-based systems with higher-dimensional spectral encoding and improved classification performance.

\backmatter

\bmhead{Acknowledgments}

Joana Teixeira and Tomás Lopes acknowledge the support of the Foundation for Science and Technology (FCT), Portugal, through Grants 2024.00426.BD and 2024.01830.BD, respectively. This project was co-financed by Component 5 - Capitalization and Business Innovation, integrated in the Resilience Dimension of the Recovery and Resiliency Plan within the scope of the Recovery and Resilience Mechanism (MRR) of the European Union (EU), framed in the Next Generation EU, for the period 2021–2026 within the scope of the HfPT project with reference 41. Tiago D. Ferreira acknowledges the support of FCT under the grant 2024.10684.CEECIND.

\section*{Data availability}
The data and code used in the production of this manuscript can be made available upon reasonable request.

\bibliography{sn-bibliography}

@article{gao2015optical,
  title={Optical hyperspectral imaging in microscopy and spectroscopy--a review of data acquisition},
  author={Gao, Liang and Smith, R Theodore},
  journal={Journal of biophotonics},
  volume={8},
  number={6},
  pages={441--456},
  year={2015},
  publisher={Wiley Online Library}
}

@article{hagen2013review,
  title={Review of snapshot spectral imaging technologies},
  author={Hagen, Nathan and Kudenov, Michael W},
  journal={Optical Engineering},
  volume={52},
  number={9},
  pages={090901--090901},
  year={2013},
  publisher={Society of Photo-Optical Instrumentation Engineers}
}

@article{liu2018novel,
  title={A novel stereo vision measurement system using both line scan camera and frame camera},
  author={Liu, Zhen and Wu, Suining and Wu, Qun and Quan, Chenggen and Ren, Yiming},
  journal={IEEE Transactions on Instrumentation and Measurement},
  volume={68},
  number={10},
  pages={3563--3575},
  year={2018},
  publisher={IEEE}
}

@article{zhou2023computer,
  title={Computer vision meets microfluidics: a label-free method for high-throughput cell analysis},
  author={Zhou, Shizheng and Chen, Bingbing and Fu, Edgar S and Yan, Hong},
  journal={Microsystems \& Nanoengineering},
  volume={9},
  number={1},
  pages={116},
  year={2023},
  publisher={Nature Publishing Group UK London}
}

@article{rane2017high,
  title={High-throughput multi-parametric imaging flow cytometry},
  author={Rane, Anandkumar S and Rutkauskaite, Justina and deMello, Andrew and Stavrakis, Stavros},
  journal={Chem},
  volume={3},
  number={4},
  pages={588--602},
  year={2017},
  publisher={Elsevier}
}

@article{holzner2018optofluidic,
  title={An optofluidic system with integrated microlens arrays for parallel imaging flow cytometry},
  author={Holzner, Gregor and Du, Ying and Cao, Xiaobao and Choo, Jaebum and deMello, Andrew J and Stavrakis, Stavros},
  journal={Lab on a Chip},
  volume={18},
  number={23},
  pages={3631--3637},
  year={2018},
  publisher={Royal Society of Chemistry}
}

@article{huang2022deep,
  title={Deep imaging flow cytometry},
  author={Huang, Kangrui and Matsumura, Hiroki and Zhao, Yaqi and Herbig, Maik and Yuan, Dan and Mineharu, Yohei and Harmon, Jeffrey and Findinier, Justin and Yamagishi, Mai and Ohnuki, Shinsuke and others},
  journal={Lab on a Chip},
  volume={22},
  number={5},
  pages={876--889},
  year={2022},
  publisher={Royal Society of Chemistry}
}

@article{siu2023optofluidic,
  title={Optofluidic imaging meets deep learning: from merging to emerging},
  author={Siu, Dickson MD and Lee, Kelvin CM and Chung, Bob MF and Wong, Justin SJ and Zheng, Guoan and Tsia, Kevin K},
  journal={Lab on a Chip},
  volume={23},
  number={5},
  pages={1011--1033},
  year={2023},
  publisher={Royal Society of Chemistry}
}

@article{lenero2018applications,
  title={Applications of event-based image sensors—Review and analysis},
  author={Le{\~n}ero-Bardallo, Juan A and Carmona-Gal{\'a}n, Ricardo and Rodr{\'\i}guez-V{\'a}zquez, Angel},
  journal={International Journal of Circuit Theory and Applications},
  volume={46},
  number={9},
  pages={1620--1630},
  year={2018},
  publisher={Wiley Online Library}
}

@article{cabriel2023event,
  title={Event-based vision sensor for fast and dense single-molecule localization microscopy},
  author={Cabriel, Cl{\'e}ment and Monfort, Tual and Specht, Christian G and Izeddin, Ignacio},
  journal={Nature Photonics},
  volume={17},
  number={12},
  pages={1105--1113},
  year={2023},
  publisher={Nature Publishing Group UK London}
}

@article{willert2022event,
  title={Event-based imaging velocimetry: an assessment of event-based cameras for the measurement of fluid flows},
  author={Willert, Christian E and Klinner, Joachim},
  journal={Experiments in Fluids},
  volume={63},
  number={6},
  pages={101},
  year={2022},
  publisher={Springer}
}

@article{howell2020high,
  title={High-speed particle detection and tracking in microfluidic devices using event-based sensing},
  author={Howell, Jessie and Hammarton, Tansy C and Altmann, Yoann and Jimenez, Melanie},
  journal={Lab on a Chip},
  volume={20},
  number={16},
  pages={3024--3035},
  year={2020},
  publisher={Royal Society of Chemistry}
}

@inproceedings{wang2020stereo,
  title={Stereo event-based particle tracking velocimetry for 3d fluid flow reconstruction},
  author={Wang, Yuanhao and Idoughi, Ramzi and Heidrich, Wolfgang},
  booktitle={European conference on computer vision},
  pages={36--53},
  year={2020},
  organization={Springer}
}

@article{banu2024hyperspectral,
  title={Hyperspectral microscopy-applications of hyperspectral imaging techniques in different fields of science: a review of recent advances},
  author={Banu, Kazi Saima and Lerma, Maricarmen and Ahmed, Sharif Uddin and Gardea-Torresdey, Jorge L},
  journal={Applied Spectroscopy Reviews},
  volume={59},
  number={7},
  pages={935--958},
  year={2024},
  publisher={Taylor \& Francis}
}

@article{zhang2023accumulation,
  title={Accumulation of nanoplastics in human cells as visualized and quantified by hyperspectral imaging with enhanced dark-field microscopy},
  author={Zhang, Hong-Jie and Zhou, Hao-Ran and Pan, Wei and Wang, Chuan and Liu, Yue-Yue and Yang, Liuyan and Tsui, Martin Tsz-Ki and Miao, Ai-Jun},
  journal={Environment International},
  volume={179},
  pages={108134},
  year={2023},
  publisher={Elsevier}
}

@article{tran2024detection,
  title={Detection and margin assessment of thyroid carcinoma with microscopic hyperspectral imaging using transformer networks},
  author={Tran, Minh Ha and Ma, Ling and Mubarak, Hasan and Gomez, Ofelia and Yu, James and Bryarly, Michelle and Fei, Baowei},
  journal={Journal of biomedical optics},
  volume={29},
  number={9},
  pages={093505--093505},
  year={2024},
  publisher={Society of Photo-Optical Instrumentation Engineers}
}

@article{chen2026self,
  title={Self-calibrated neuromorphic hyperspectral derivative imaging},
  author={Chen, Rongzhou and Wang, Chutian and Li, Yuxing and Cao, Yuqing and Zhu, Shuo and Lam, Edmund Y},
  journal={Optica},
  volume={13},
  number={4},
  pages={587--590},
  year={2026},
  publisher={Optica Publishing Group}
}

@article{franceschelli2025assessment,
  title={An assessment of event-based imaging velocimetry for efficient estimation of low-dimensional coordinates in turbulent flows},
  author={Franceschelli, Luca and Willert, Christian E and Raiola, Marco and Discetti, Stefano},
  journal={Experimental Thermal and Fluid Science},
  volume={164},
  pages={111425},
  year={2025},
  publisher={Elsevier}
}

@article{tsilikas2024photonic,
  title={Photonic neuromorphic accelerators for event-based imaging flow cytometry},
  author={Tsilikas, Ioannis and Tsirigotis, Aris and Sarantoglou, George and Deligiannidis, Stavros and Bogris, Adonis and Posch, Christoph and Van den Branden, Gerd and Mesaritakis, Charis},
  journal={Scientific Reports},
  volume={14},
  number={1},
  pages={24179},
  year={2024},
  publisher={Nature Publishing Group UK London}
}

@inproceedings{zhang2022work,
  title={Work in progress: Neuromorphic cytometry, high-throughput event-based flow flow-imaging},
  author={Zhang, Ziyao and Ma, Maria Sabrina and Eshraghian, Jason K and Vigolo, Daniele and Yong, Ken-Tye and Kavehei, Omid},
  booktitle={2022 8th International Conference on Event-Based Control, Communication, and Signal Processing (EBCCSP)},
  pages={1--5},
  year={2022},
  organization={IEEE}
}

@article{Lombardo2021,
   author = {Jeremy A. Lombardo and Marzieh Aliaghaei and Quy H. Nguyen and Kai Kessenbrock and Jered B. Haun},
   doi = {10.1038/s41467-021-23238-1},
   issn = {20411723},
   issue = {1},
   journal = {Nature Communications},
   month = {12},
   pmid = {34001902},
   publisher = {Nature Research},
   title = {Microfluidic platform accelerates tissue processing into single cells for molecular analysis and primary culture models},
   volume = {12},
   year = {2021}
}

@article{Zhang2024,
   author = {Han Zhang and Rohit Gupte and Yuwen Li and Can Huang and Adrian R. Guzman and Jeong Jae Han and Haemin Jung and Rushant Sabnis and Paul de Figueiredo and Arum Han},
   doi = {10.1038/s41467-024-52932-z},
   issn = {20411723},
   issue = {1},
   journal = {Nature Communications },
   month = {12},
   pmid = {39487108},
   publisher = {Nature Research},
   title = {NOVAsort for error-free droplet microfluidics},
   volume = {15},
   year = {2024}
}

@article{Mohammadimehr2024,
   author = {Amir Mohammadimehr and Angeles Ivón Rodríguez-Villarreal and Joan Antoni López Martínez and Jasmina Casals-Terré},
   doi = {10.1016/j.jafr.2024.101124},
   issn = {26661543},
   journal = {Journal of Agriculture and Food Research},
   month = {6},
   publisher = {Elsevier B.V.},
   title = {Review: Impact of microfluidic cell and particle separation techniques on microplastic removal strategies},
   volume = {16},
   year = {2024}
}

@misc{Khan2026,
   author = {Muhammad Soban Khan and Raihan Hadi Julio and Mushtaq Ali and Sebastian Sachs and Christian Cierpka and Jörg König and Jinsoo Park},
   doi = {10.1039/d5lc00826c},
   issn = {14730189},
   journal = {Lab on a Chip},
   month = {3},
   publisher = {Royal Society of Chemistry},
   title = {Microfluidic shape-based separation for cells and particles: recent progress and future perspective},
   year = {2026}
}

\end{document}